# Development of a 4D Cerebral Microvascular Imaging Platform for Mouse Stroke Model

Yoshihisa Kaneko
*Graduate School of Biomedical Engineering*
*Tohoku university*
Sendai, Japan
kaneko.yoshihisa.q1@dc.tohoku.ac.jp

Moe Kumai
*Graduate School of Medicine*
*Tohoku University*
Sendai, Japan
moe.kumai.s2@dc.tohoku.ac.jp

Hiroyuki Igarashi
*Graduate School of Pharmaceutical Sciences*
*Tohoku university*
Sendai, Japan
hiroyuki.igarashi.e3@tohoku.ac.jp

Daisuke Ando
*Graduate School of Medicine*
*Tohoku University*
Sendai, Japan
daisuke.ando.a5@tohoku.ac.jp

Kuniyasu Niizuma
*Graduate School of Medicine*
*Tohoku University*
Sendai, Japan
kniizuma@tohoku.ac.jp

Hidenori Endo
*Graduate School of Medicine*
*Tohoku University*
Sendai, Japan
h-endo@tohoku.ac.jp

Yoshifumi Saijo
*Graduate School of Biomedical Engineering*
*Tohoku university*
Sendai, Japan
saijo@tohoku.ac.jp

Takuro Ishii
*Graduate School of Biomedical Engineering*
*Tohoku university*
Sendai, Japan
takuro.ishii@tohoku.ac.jp

***Abstract*— In ischemic stroke, changes in cerebral hemodynamics during both the ischemic and reperfusion phases strongly influence stroke outcomes. However, these hemodynamic changes remain incompletely understood. To address this challenge, we devised an imaging platform that enables time-resolved ultrasound microvascular imaging during the experimental induction of ischemia and reperfusion in a mouse model. The platform leverages our previous ultrasound imaging framework combined with continuous mechanical scanning, which acquires whole-brain blood-flow signals within 5 s. The experiments demonstrated that the proposed platform can visualize both local and whole-brain hemodynamic responses to the induction of ischemia and reperfusion, suggesting its potential for rapid and continuous whole-brain hemodynamic assessment in small-animal models.**

***Keywords—ultrafast ultrasound, Doppler imaging, micro-vasculature imaging, whole-brain, stroke, tMCAO, microbubble***

## I. Introduction

Stroke is the third-leading cause of death and disability combined worldwide, imposing a substantial global health burden [1]. In ischemic stroke, changes in cerebral blood flow are not confined to the ischemic territory but can extend across the brain, partly because cerebral autoregulation may be impaired [2-3]. Cerebral hemodynamics during ischemia and subsequent reperfusion are associated with tissue viability, such as that of the ischemic penumbra, and with stroke outcomes [4-6]. Nevertheless, these hemodynamic alterations, particularly at the whole-brain level, remain incompletely understood [7-8]. Therefore, investigating how cerebral blood flow evolves from arterial occlusion through reperfusion using experimentally controlled animal models is important.

Various imaging modalities have been used to investigate cerebral hemodynamics. For example, functional magnetic resonance imaging (fMRI) and functional ultrasound (fUS) provide broad hemodynamic coverage but limited vessel-scale visualization [9-10]. In contrast, two-photon microscopy offers high-resolution vascular imaging but limited field of view and imaging depth [11-12]. Thus, an imaging approach that combines brain-wide coverage, microvascular resolvability, and sufficiently rapid volumetric acquisition would be valuable for investigating dynamic cerebrovascular responses.

In this study, we aimed to develop a four-dimensional microvascular imaging platform by extending our previous rapid three-dimensional imaging framework [13], [14] to enable ultrasound imaging during surgical induction of ischemia and controlled reperfusion in small-animal models. We evaluated the platform by capturing whole-brain hemodynamic responses during ischemia and reperfusion in a mouse stroke model.

## II. Material and Methods

### A. Animal Preparation for Imaging

To experimentally induce cerebral ischemia and reperfusion, a transient middle cerebral artery occlusion (tMCAO) model mouse was used. One side of the middle cerebral artery (MCA) was transiently occluded by inserting an occluding monofilament from the internal carotid artery (ICA) toward the MCA origin at the Circle of Willis (Fig. 1)[15], [16]. After a defined ischemic period (60 min), the filament was withdrawn to induce reperfusion.

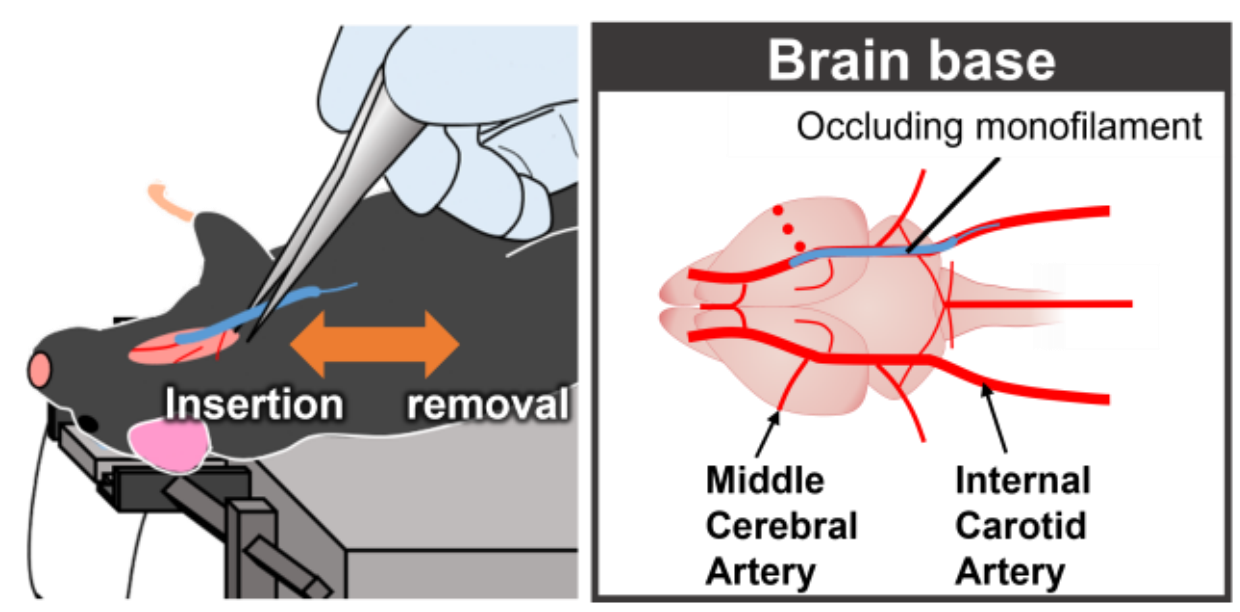

Fig. 1. Schematic illustration of the tMCAO procedure. An occluding monofilament is inserted through the ICA to induce cerebral ischemia. The monofilament is withdrawn to initiate reperfusion.

This work was supported by the Creative Interdisciplinary Collaboration Program at the Frontier Research Institute for Interdisciplinary Sciences, Tohoku University, and AMED under Grant Number JP25ym0126818.

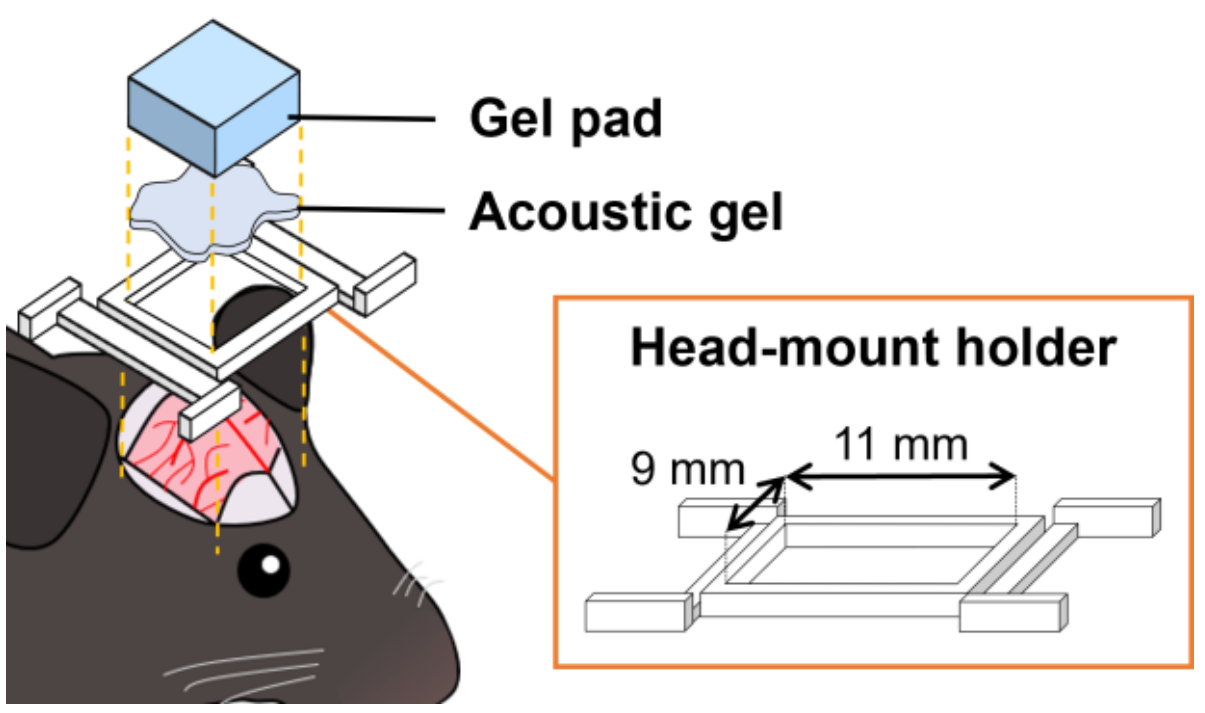

Fig. 2. Schematic illustration of the acoustic window and custom-made head-mount holder.

A 9-week-old C57BL/6 mouse was prepared for ultrasound imaging during the tMCAO procedure. Dexamethasone was administered to suppress cerebral edema, followed by craniotomy to provide an acoustic window. A custom-made head-mount holder with a 9 × 11 mm square opening was then fixed to the head, and a gel pad (HYD110, HydroAid, Kikgel, Poland) and acoustic gel were placed within the opening to provide acoustic coupling between the ultrasound transducer and the exposed brain surface (Fig. 2). After the preparation, the mouse was transferred to the imaging setup described in the next section.

All animal preparation and data acquisition were performed under isoflurane anesthesia. The depth of anesthesia was continuously monitored and adjusted by a neurosurgeon throughout the experiment. A heating pad was used to maintain the animal's body temperature. This study was conducted with the approval of the Animal Experiment Ethics Committee of Tohoku University (approval number: 2020BeA-009-04) and in accordance with the National Institutes of Health (NIH) guidelines for the care and use of laboratory animals.

### *B. Imaging System Setup and Imaging Sequence*

To enable ultrasound imaging during the surgical induction of ischemia and reperfusion, we devised an imaging setup that provides simultaneous access to the brain for imaging and to the ventral neck for surgical manipulation. Since the tMCAO procedure requires surgical access to the carotid arteries through the ventral neck, the mouse was fixed in the supine position by securing the aforementioned head-mount holder to a stereotaxic fixation device (MAG-1 + MAG-A, Narishige, Japan). The ultrasound transducer was positioned underneath the head and applied to the cranial acoustic window through the gel pad and acoustic gel.

An overview of the imaging setup is shown in Fig. 3. Ultrasound imaging was performed using a research-purpose ultrasound platform (Vantage 256 high-frequency configuration, Verasonics Inc., USA) equipped with an L38-22v high-frequency linear array transducer ($f_c$: 30 MHz; elevational focus: 8.0 mm; active aperture: 128 elements, 8.8 mm; KOLO Silicon, USA). The transducer was mounted on a mechanical translation stage (OSMS(CS)20-35, SIGMAKOKI Co., Ltd., Japan) and continuously scanned along the elevational direction at 1 mm/s for 5 s.

The start of the translational motion and the ultrasound transmit-receive sequence were synchronized using a trigger pulse generated by an Arduino microcontroller so that the predefined imaging sequence could acquire a dataset over a volume of 7 × 8 × 5 mm in the axial, lateral, and elevational directions, respectively.

The imaging sequence was programmed to acquire plane wave compounded frames consisting of ten raw frames with steering angles spanning ±5° at a pulse repetition frequency (PRF) of 5 kHz, resulting in a post-compounding frame rate of 500 Hz. During the 5-s mechanical scan, a total of 2,500 compounded B-mode frames were acquired. Each compounded frame was acquired in 2 ms, while the transducer translated by 2 µm.

### *C. Data Acquisition Protocol*

Three-dimensional ultrasound imaging was sequentially performed at eight time points during the tMCAO procedure (Fig. 4): before ischemia (control), at 1, 15, and 60 min after MCA occlusion, and at 1, 15, 20 and 30 min after reperfusion. After the 60-min ischemic acquisition, the occluding filament was withdrawn to initiate reperfusion. One minute before each acquisition, 100 µL of microbubble suspension was injected through the contralateral ICA as an ultrasound contrast agent [17].

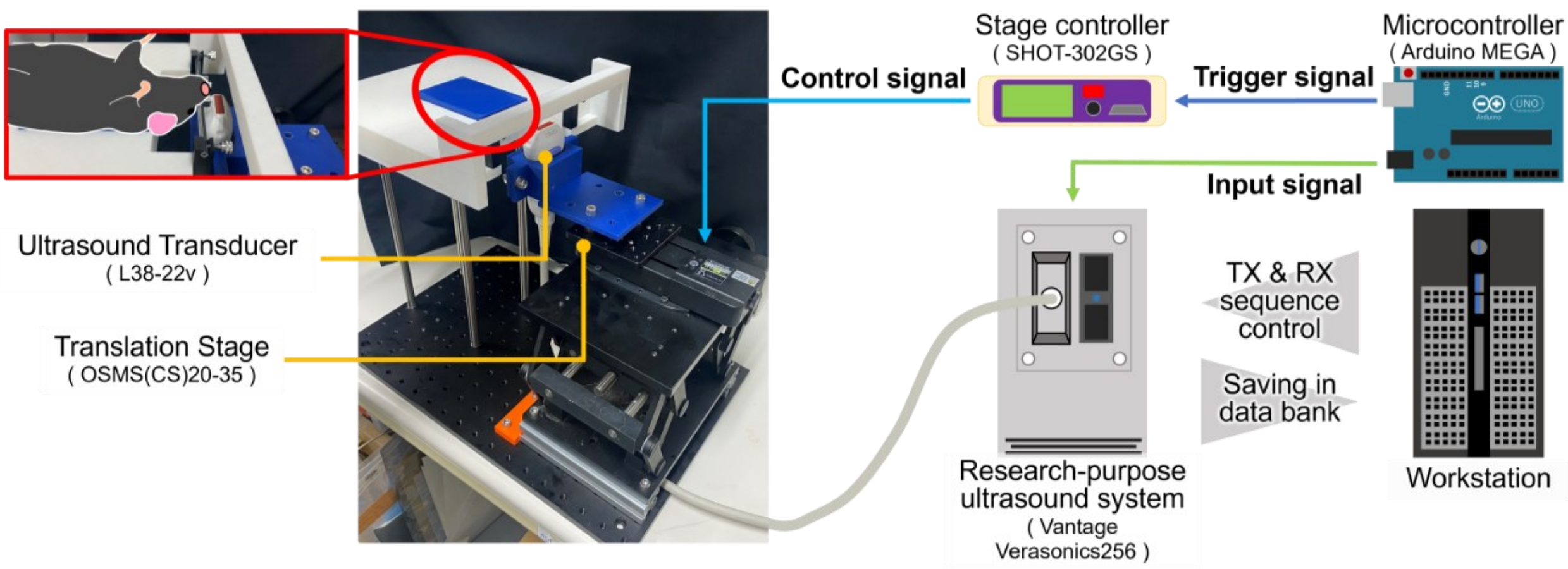

Fig. 3. Overview of the imaging system setup. The mouse was placed in the supine position on the imaging bed, and the ultrasound transducer was applied from underneath the head.

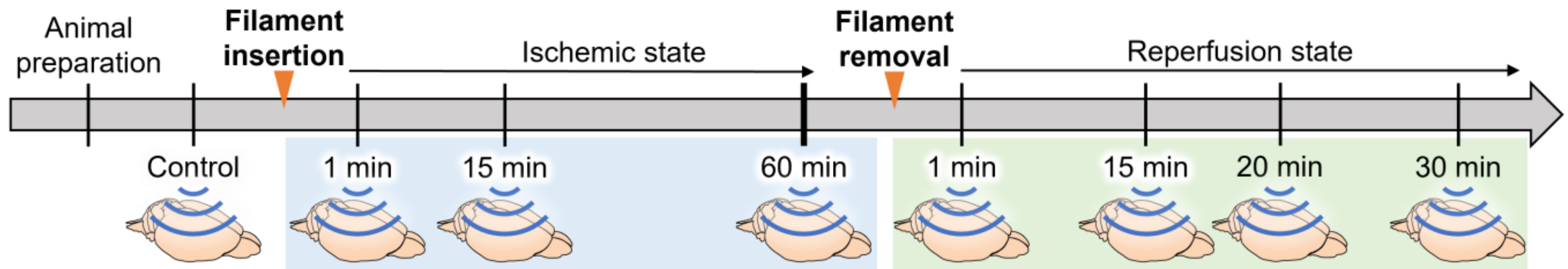

Fig. 4. Timeline of the tMCAO procedure and ultrasound imaging. Cerebral ischemia was induced by occluding the middle cerebral artery for 60 min. Ultrasound datasets were acquired at eight time points: one before ischemia (Control), three during ischemia (1, 15 and 60 min), and four after reperfusion (1, 15, 20 and 30 min).

### *D. Signal and Image Processing Pipeline*

Following delay-and-sum beamforming and coherent compounding, the acquired datasets were processed using our previously reported section-wise singular value decomposition (SVD) framework[13], [14] in order to reconstruct three-dimensional microvascular images. For each volumetric acquisition, the 2,500 compounded frames were divided into 49 sections, each consisting of 100 frames with a 50-frame overlap (Fig. 5). SVD filtering was then applied independently to each section to extract blood-flow signals.

To maintain consistent blood-flow extraction within each volume, a common pair of SVD cutoff indices, i.e., $p$ and $q$ in Fig. 5, was determined from four representative sections (sections 10, 20, 30, and 40). The cutoff indices were selected to encompass the blood-flow components identified across these representative sections and were then applied to all 49 sections within the volume. This threshold selection procedure was performed independently for each of the eight volumetric acquisitions. Power Doppler images were then generated from the extracted blood-flow signals and arranged along the elevational direction to reconstruct the three-dimensional microvascular map.

Finally, the time-series 3D vascular images were spatially aligned using 3D registration with mutual information as the similarity metric, enabling visualization and quantification of hemodynamic changes in the same vessels.

## III. RESULTS AND DISCUSSION

The proposed platform successfully enabled four-dimensional cerebral microvascular imaging throughout the tMCAO procedure without repositioning the mouse, while maintaining surgical access to the ventral neck for induction of ischemia and reperfusion. Fig. 6 shows the imaging results obtained at all eight time points. The images are maximum intensity projections (MIPs) of the elevational sections covering the MCA territory.

Compared with the control condition, blood-flow signals decreased in both cortical and deep regions during MCA occlusion. In the right cortical region within the MCA territory, the blood-flow signal intensity decreased by 6.7 dB during ischemia and subsequently increased by 12.2 dB after reperfusion. Blood-flow signals also reappeared in the deep region following filament withdrawal. These results indicate that the proposed platform captured both spatial and temporal hemodynamic responses associated with ischemia and reperfusion while the animal remained fixed in the imaging setup. From approximately 20 min after reperfusion, the physiological condition of the mouse deteriorated. Therefore, the reduced blood-flow signals observed at 20 and 30 min after the event may partly reflect systemic physiological deterioration.

The platform also retained the whole-brain-scale imaging capability of our previously reported framework [14], allowing visualization of vascular signals from the cortical surface to deep and basal brain regions. Combined with the fixed-animal configuration, this wide spatial coverage enables repeated assessment of the same brain during ischemia and reperfusion without repositioning.

In the present experiment, the interval between volumetric acquisitions ranged from approximately 1 min to 45 min depending on the experimental protocol. However, each volumetric acquisition required only 5 s. Even when the return motion of the translation stage is considered, repeated volumetric imaging at intervals as short as 10 s would be technically feasible, potentially allowing more rapid hemodynamic responses to the pathophysiological changes to be quantified. Thus, the temporal sampling used in this study was determined mainly by experimental protocol rather than by the acquisition speed of the imaging platform. Further development is required for high-frequency repeated imaging. In addition, the present study was a feasibility demonstration using a single tMCAO mouse, and further studies with multiple animals will be required to evaluate reproducibility and quantitative hemodynamic changes during ischemia and reperfusion.

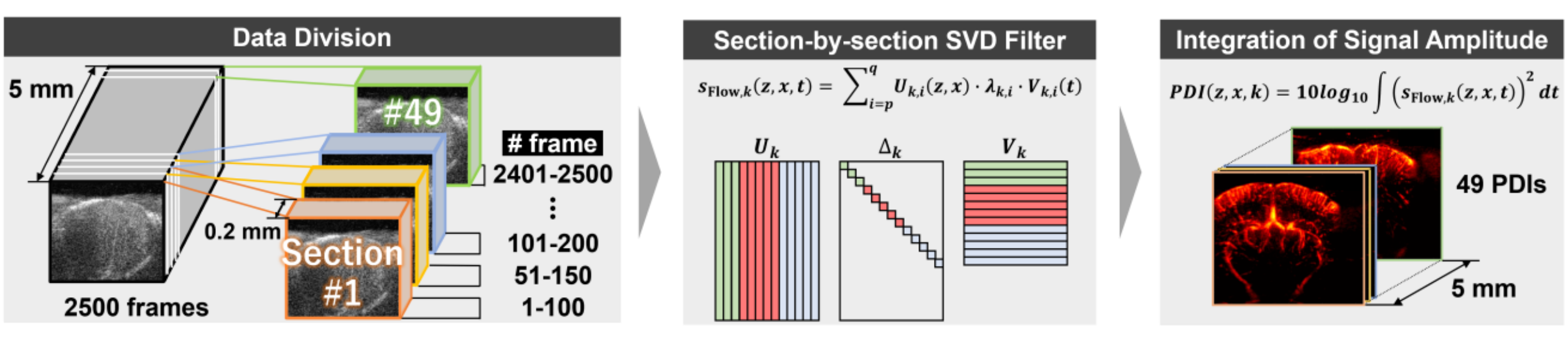

Fig. 5. Processing pipeline of the section-wise SVD filter.

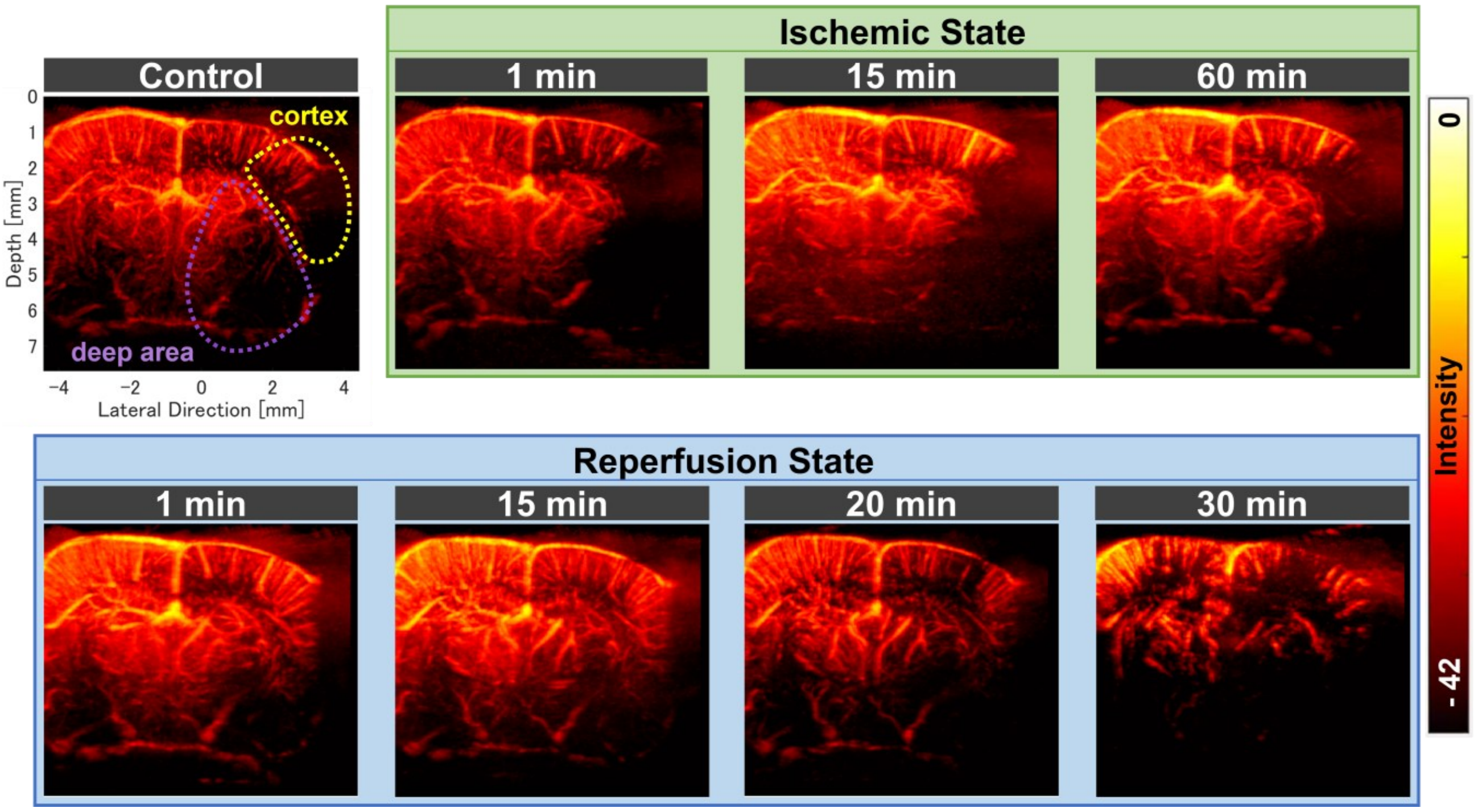

Fig. 6. Imaging results obtained at all eight time points. Within the MCA territory, blood-flow signals decreased in both cortical and deep regions during ischemia and reappeared following reperfusion.

## IV. Conclusion

In this study, we developed an imaging platform that enables four-dimensional whole-brain microvascular imaging during the tMCAO surgical procedure without repositioning the animal. The experiment demonstrated the feasibility of the proposed platform for rapid and continuous assessment of cerebral hemodynamics during experimentally induced ischemia and reperfusion.